\documentclass[pre,twocolumn,groupedaddress,showpacs,showkeys,amsmath,amssymb,floatfix]{revtex4-1}

\usepackage{graphicx}
\usepackage{dcolumn}
\usepackage{bm}
\usepackage{multirow}

\usepackage{subfigure}

\graphicspath{{images/}{plots/}}

\newcommand{\brac}[1]{\ensuremath{\left(#1\right)}}
\newcommand{\duck}[1]{\ensuremath{\left<#1\right>}}

\newcommand{\deff}{{\ensuremath{d_{\mathrm{e}}}}}

\renewcommand{\vec}[1]{\bm{#1}}

\DeclareMathOperator{\ee}{e}

\DeclareMathOperator{\dist}{dist}

\begin{document}

    \title{Large Deviations of Convex Hulls of Self-Avoiding Random Walks}
    \author{Hendrik Schawe}
    \email{hendrik.schawe@uni-oldenburg.de}
    \affiliation{Institut f\"ur Physik, Universit\"at Oldenburg, 26111 Oldenburg, Germany}
    \affiliation{LPTMS, CNRS, Univ.~Paris-Sud, Universit\'e Paris-Saclay, 91405 Orsay, France}
    \author{Alexander K. Hartmann}
    \email{a.hartmann@uni-oldenburg.de}
    \affiliation{Institut f\"ur Physik, Universit\"at Oldenburg, 26111 Oldenburg, Germany}
    \affiliation{LPTMS, CNRS, Univ.~Paris-Sud, Universit\'e Paris-Saclay, 91405 Orsay, France}
    \author{Satya N. Majumdar}
    \email{satya.majumdar@u-psud.fr}
    \affiliation{LPTMS, CNRS, Univ.~Paris-Sud, Universit\'e Paris-Saclay, 91405 Orsay, France}
    \date{\today}

    \begin{abstract}
        A global picture of a random particle movement
        is given by the convex hull of the visited points.
        We obtained  numerically the probability
         distributions of the volume and surface of
        the convex hulls of a selection of three types of self-avoiding
        random walks, namely the classical Self-Avoiding Walk,
        the Smart-Kinetic Self-Avoiding Walk, and the Loop-Erased
        Random Walk. To obtain a comprehensive description of the
        measured random quantities, we applied sophisticated large-deviation
        techniques, which allowed us to
        obtain the distributions over a large range of the support
        down to probabilities far smaller than $P = 10^{-100}$.
        We give an approximate closed form of the so-called large-deviation
        rate function $\Phi$ which generalizes
        above the upper critical dimension to the previously studied
        case of the standard random walk. Further we show
        correlations between the two observables also in the limits of
        atypical large or small values.
    \end{abstract}

    \pacs{02.50.-r, 75.40.Mg, 89.75.Da}

    \maketitle

    \section{Introduction}
        The standard random walk is a simple Markovian process, which
        has a history as a model for diffusion. There are many exact results
        known \cite{hughes1996random}. If memory is added to the model, e.g.,
        to interact with the past trajectory of the walk, analytic treatment
        becomes much harder. A class of self-interacting random walks that we will
        focus on in this study, are \emph{self-avoiding} random walks, which
        live on a lattice and do not visit any site twice. This can be used to
        model systems with excluded volume, e.g., polymers whose single monomers
        can not occupy the same site at once \cite{Madras2013}.
        There are more applications which are not as obvious, e.g., a slight
        modification of the \emph{Smart-Kinetic Self-Avoiding Walk} traces
        the perimeter of critical percolation clusters \cite{weinrib1985kinetic},
        while the \emph{Loop-Erased Random Walk} can be used to study
        spanning trees \cite{manna1992spanning} (and vice versa \cite{Majumdar1992Exact}).

        One of the central properties of random walk models is the exponent $\nu$,
        which characterizes the growth of the end-to-end distance $r$ with the
        number of steps $T$, i.e., $r \propto T^\nu$. While this has the value
        $\nu = 1/2$ for the standard random walk, its value is larger for the
        self-avoiding variations, which are effectively pushed away from their
        past trajectory. In two dimensions, this value (and other properties)
        can often be obtained by the correspondence to
        Schramm-Loewner evolution \cite{cardy2005sle,lawler2002scaling,Lawler2011,Kennedy2015}.
        But between two dimensions and the upper critical dimension,
        above which the
        behavior is the same as the standard random walk, Monte Carlo
        simulations are used to obtain estimates for the exponent $\nu$.

        Here we want to study the convex hulls of a selection of self-avoiding
        walk models featuring larger values of $\nu$. The convex
        hull allows one to obtain a global picture of the space occupied
        by a walk, without exposing all details of the walk. As an example,
        convex hulls are used to describe the home ranges of animals
        \cite{mohr1947,worton1987,boyle2009}. Namely, we will look at
        the \emph{Smart-Kinetic Self-Avoiding Walk} (SKSAW), the classical
        \emph{Self-Avoiding Walk} (SAW) and the
        \emph{Loop-Erased Random Walk} (LERW), since they span a large range
        of $\nu$ values and are well established in the literature.
        About the convex hulls of standard random walks we already know plenty
        properties. The mean perimeter and area are known exactly since over
        20 years \cite{Letac1980Expected,Letac1993explicit}
        for large walk lengths $T$, i.e., the Brownian Motion limit.
        Since then simpler and more general methods were devised,
        which are based on using Cauchy's formula with relates
        the support function of a curve to the perimeter and the area
        enclosed by the curve \cite{Majumdar2009Convex,Majumdar2010Random}.
        More recently also the mean hypervolume and surface for arbitrary
        dimensions was calculated \cite{Eldan2014Volumetric}.
        For discrete-time random walks with jumps from an arbitrary distribution,
        the perimeter of the convex hull for finite (but large) walk lengths $T$
        were computed explicitly \cite{grebenkov2017mean}. For the case of Gaussian jump lengths even an
        exact combinatorial formula for the volume in arbitrary dimensions is known \cite{kabluchko2016intrinsic}.
        For the variance there is an exact result for Brownian bridges
        \cite{Goldman1996}. Concerning the full distributions,
        no exact analytical results are available. Here sophisticated
        large-deviation
        simulations were used to numerically explore a large part of the full
        distribution, i.e., down to probabilities far smaller than $10^{-100}$ \cite{Claussen2015Convex,Dewenter2016Convex,schawe2017highdim}.

        Despite this increasing interest in the convex hulls of standard random
        walks, there seem to be no studies treating the convex hulls of
        self-avoiding walks.
        To fill this void, we use Markov chain Monte Carlo sampling to obtain
        the distributions of some quantities of interest over their whole support.
        To connect to previous studies \cite{Claussen2015Convex,Dewenter2016Convex,schawe2017highdim}
        we also compare the aforementioned variants to the standard random
        walk on a square lattice (LRW). We are mainly
        interested in the full distribution of the area $A$ and the perimeter
        $L$ of $d=2$ dimensional hulls for walks in the plane, since the
        effects of the self-interactions are stronger in lower dimensions.
        Though, we will also look into the volume $V$ in the $d=3$
        dimensional case. In the past study on standard random walks \cite{schawe2017highdim}
        we found that the full distribution can be
        scaled to a universal distribution using only the exponent $\nu$ and
        the dimension for large walk lengths $T$.
        For the present case, where a walk might depend on its full history,
        one could expect a more complex behavior. Nevertheless, our results
        presented below show convincingly that also for self-interacting walks the
        distributions are universal and governed mainly by the exponent $\nu$,
        except for some finite-size effects, which are probably caused by the
        lattice structure.
        Further we use the distributions to obtain empirical large-deviation
        rate functions \cite{Touchette2009large}, which suggests that a limiting
        rate function is mathematically well defined. We also give an estimate
        for the rate function, which is compatible with the known case of
        standard random walks and with all cases under scrutiny in this study.

    \section{Models and Methods}\label{sec:mm}
        This sections gives a short overview over the used models and methods,
        with references to literature more specialized on the corresponding
        subject. Where we deem adequate, also technical details applicable for
        this study are mentioned.

    \subsection{Sampling Scheme}
\begin{figure}[bhtp]
            \centering
            \includegraphics[width=0.7\linewidth]{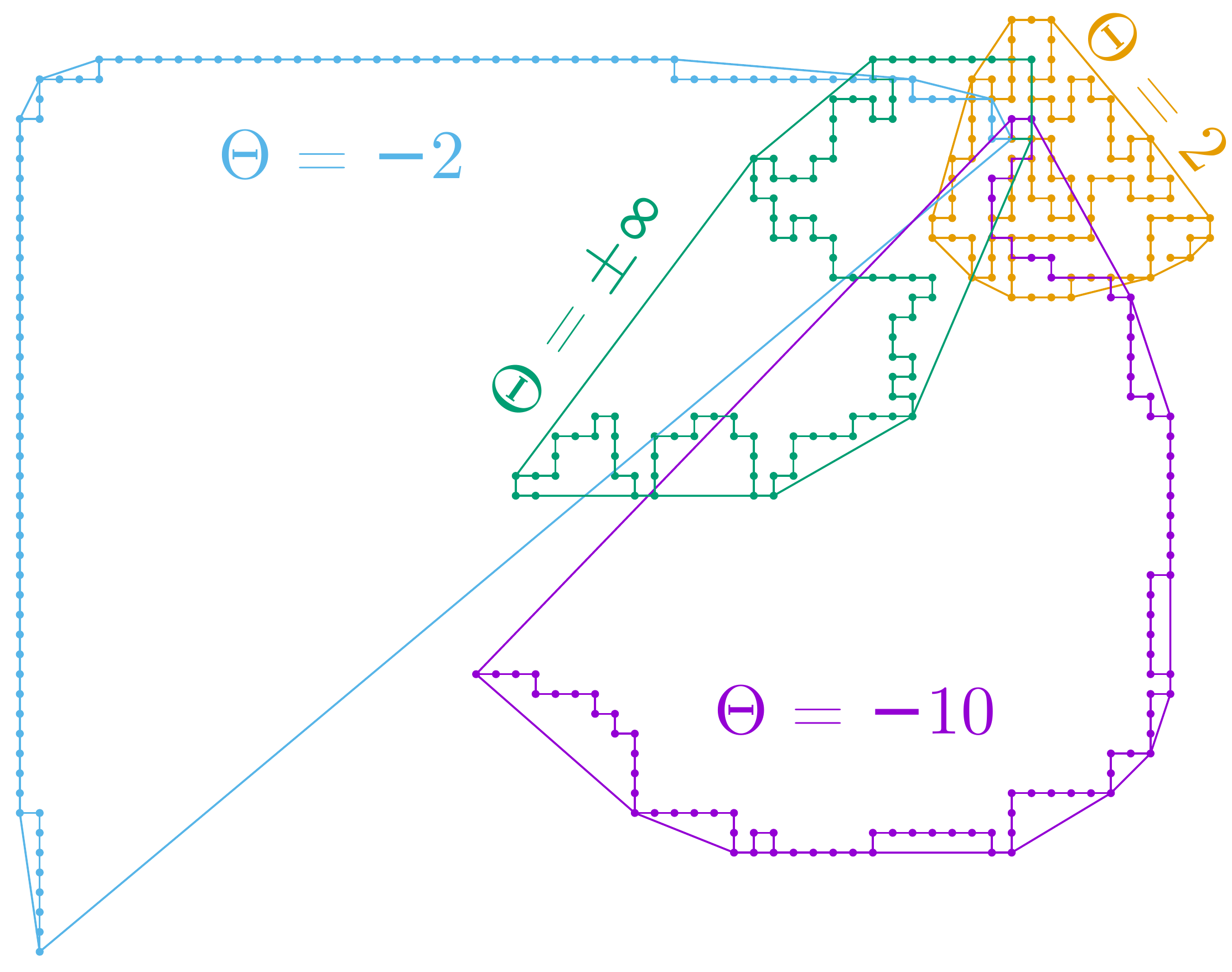}
            \caption{\label{fig:saw_temp_cmp}
                (color online)
                Typical SAW configurations with
                $T=200$ steps and their convex hulls at different temperatures $\Theta$.
                $\Theta = \pm \infty$ corresponds to a typical configuration
                without bias.
            }
        \end{figure}

        To generate the whole distribution of the area or perimeter of the convex
        hull of a random walk over its full support, a
        sophisticated Markov chain Monte Carlo (MCMC) sampling scheme is
        applied \cite{Hartmann2002Sampling,Hartmann2011}. The Markov chain is
        here a sequence of different walk configurations.
        The fundamental idea is to treat the observable $S$, i.e., the
        perimeter, area or volume, as the energy of a physical system which is
        coupled to a heat bath with adjustable ``temperature'' $\Theta$
        and to sample its equilibrium distribution using the Markov chain. This
        can be easily done using the classical Metropolis algorithm
        \cite{metropolis1953equation}.
        Therefore the current walk configuration is changed a bit (the precise
        type of change is dependent on the type of walk, we are looking at and
        is explained in the following sections). The change is accepted with
        the acceptance probability
        \begin{align}
            \label{eq:pacc}
            p_\mathrm{acc} = \min\{1,\ee^{-\Delta S / \Theta}\}
        \end{align}
        and rejected otherwise.
        The $\Theta$ will then bias the configuration towards specific ranges
        of $S$. Configurations at small and negative $\Theta$ will show larger
        than typical $S$, small and positive $\Theta$ show smaller than typical
        $S$ and large values independent of the sign sample configurations from
        the peak of the distribution. Fig.~\ref{fig:saw_temp_cmp} shows typical
        walk configurations of the self-avoiding walk at different
        values of $\Theta$.

        In a second step, histograms of the equilibrium distribution
        $P_\Theta(S)$ are corrected for the bias introduced via $\Theta$
        as follows.
        \begin{align}
            P(S) = \ee^{S / \Theta} Z(\Theta) P_\Theta(S)
        \end{align}
        The free parameter $Z(\Theta)$ can be obtained by enforcing continuity
        and normalization of the distribution. We do not present
        further details here, because the algorithm \cite{Hartmann2002Sampling}
        has been applied and explained in detail several times, also in a very
        general form \cite{Hartmann2014high}. In particular, the
        algorithm was already used successfully in other studies looking
        at the large deviation properties of convex hulls of random walks
        \cite{Claussen2015Convex,Dewenter2016Convex}.

    \subsection{Lattice Random Walk (LRW)}
        All of the self-interacting random walks, which are the focus of this
        study, are typically treated on a lattice. Hence, we will start by
        introducing the simple, i.e., non-interacting, isotropic random walk on a lattice.
        For simplicity we will use a square lattice with a lattice constant of
        $1$. A realization consists of $T$ randomly chosen discrete steps
        $\vec\delta_i$. Here we use steps between adjacent lattice sites, i.e.,
        $d$-dimensional Cartesian base vectors $\vec{e}_i$,
        which are drawn uniformly from $\{\pm\vec{e}_i\}$. The realization can
        be defined as the tuple of the steps $(\vec\delta_1, ..., \vec\delta_T)$ and
        the position at time $\tau$ as
        \begin{align}
            \vec{x}(\tau) = \vec{x}_0 + \sum_{i=1}^\tau \vec{\delta}_i.
        \end{align}
        Here we set the start point $\vec{x}_0$ at the coordinate origin.
        The set of visited sites is therefore $\mathcal P = \{\vec{x}(0), ..., \vec{x}(T)\}.$

        The central quantity of the LRW is the average end-to-end distance
        \begin{equation}
            r = \sqrt{\langle (\vec x (T)- \vec{x}_0)^2 \rangle} \,,
        \end{equation}
        where $\langle \ldots \rangle$ denotes the average over the disorder.
        It grows polynomially and is characterized by the exponent $\nu$ via
        $r \propto T^\nu$. For the LRW it is $\nu = 1/2$, which
        is typical for all diffusive processes.

        As the change move for the Metropolis algorithm, we replace a randomly
        chosen $\vec\delta_i$ by a new randomly drawn displacement. Since our
        quantity of interest is the convex hull, i.e., a global property of the
        walk, we do not profit much from local moves, e.g., crankshaft moves.
        Thus we use this simple, global move.

        \begin{figure}[bhtp]
            \centering
            \subfigure[\label{fig:LRW} LRW]{
                \includegraphics[scale=0.17]{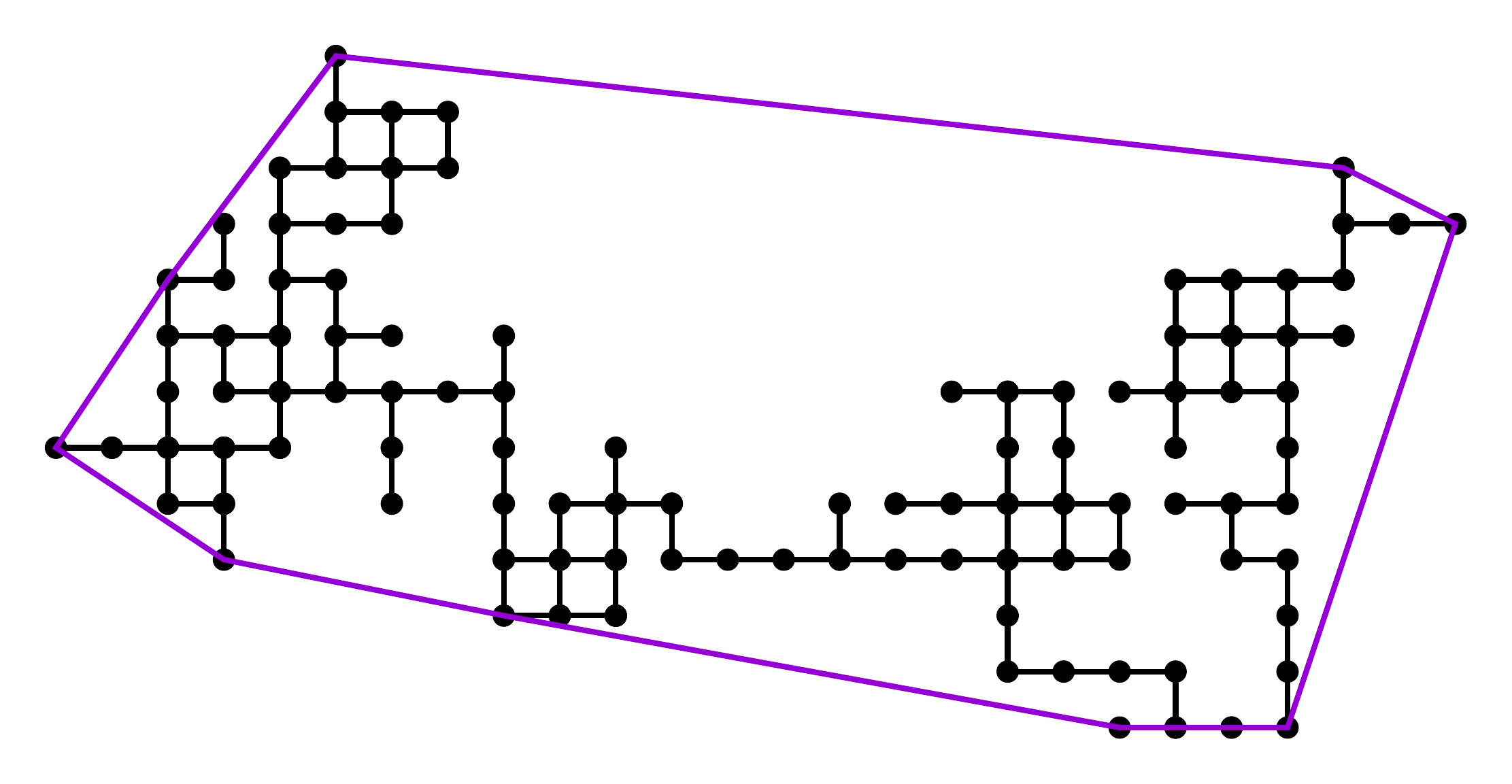}
            }
            \subfigure[\label{fig:SKSAW} SKSAW]{
                \includegraphics[scale=0.17]{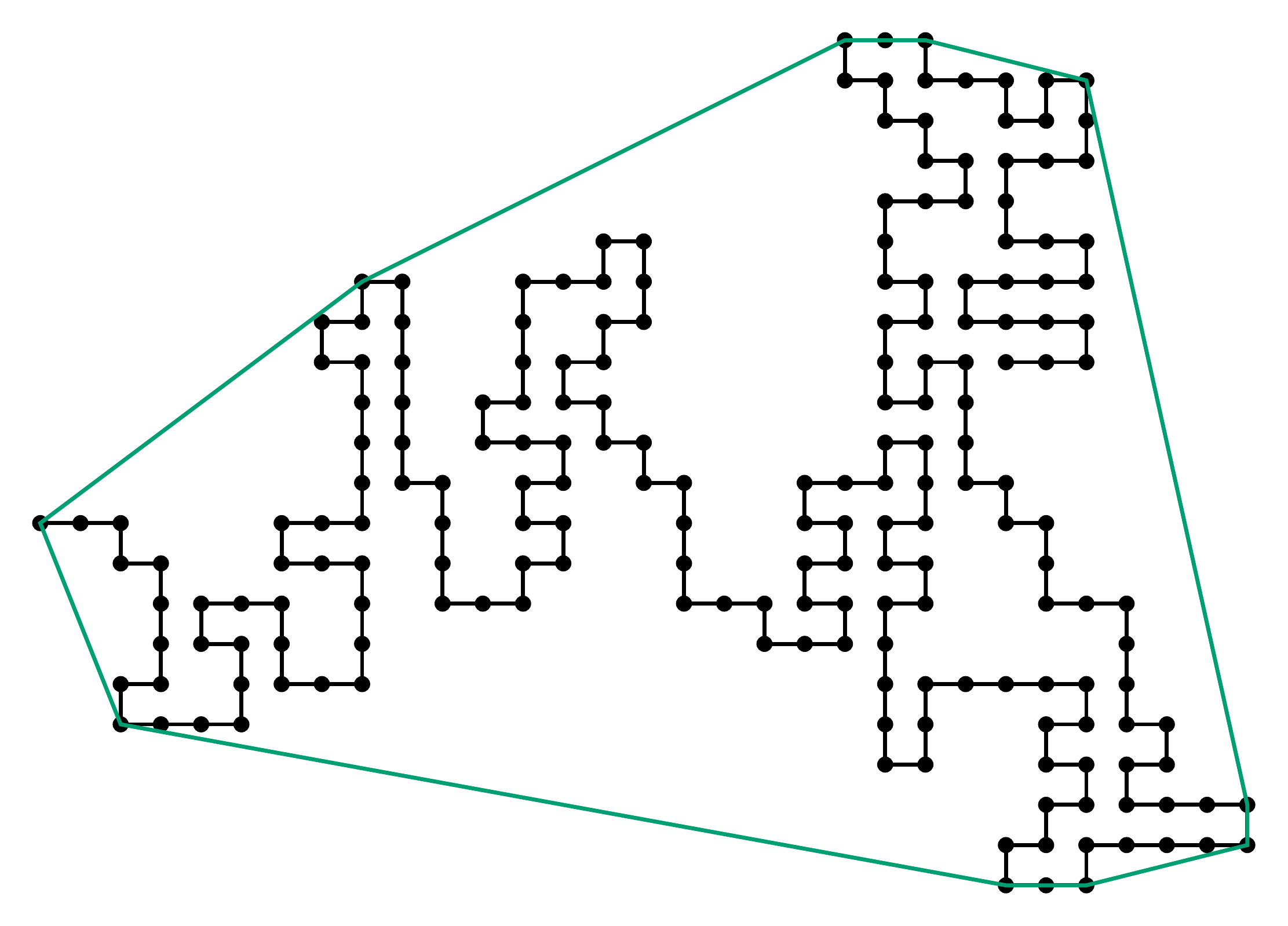}
            }

            \subfigure[\label{fig:SAW} SAW]{
                \includegraphics[scale=0.17]{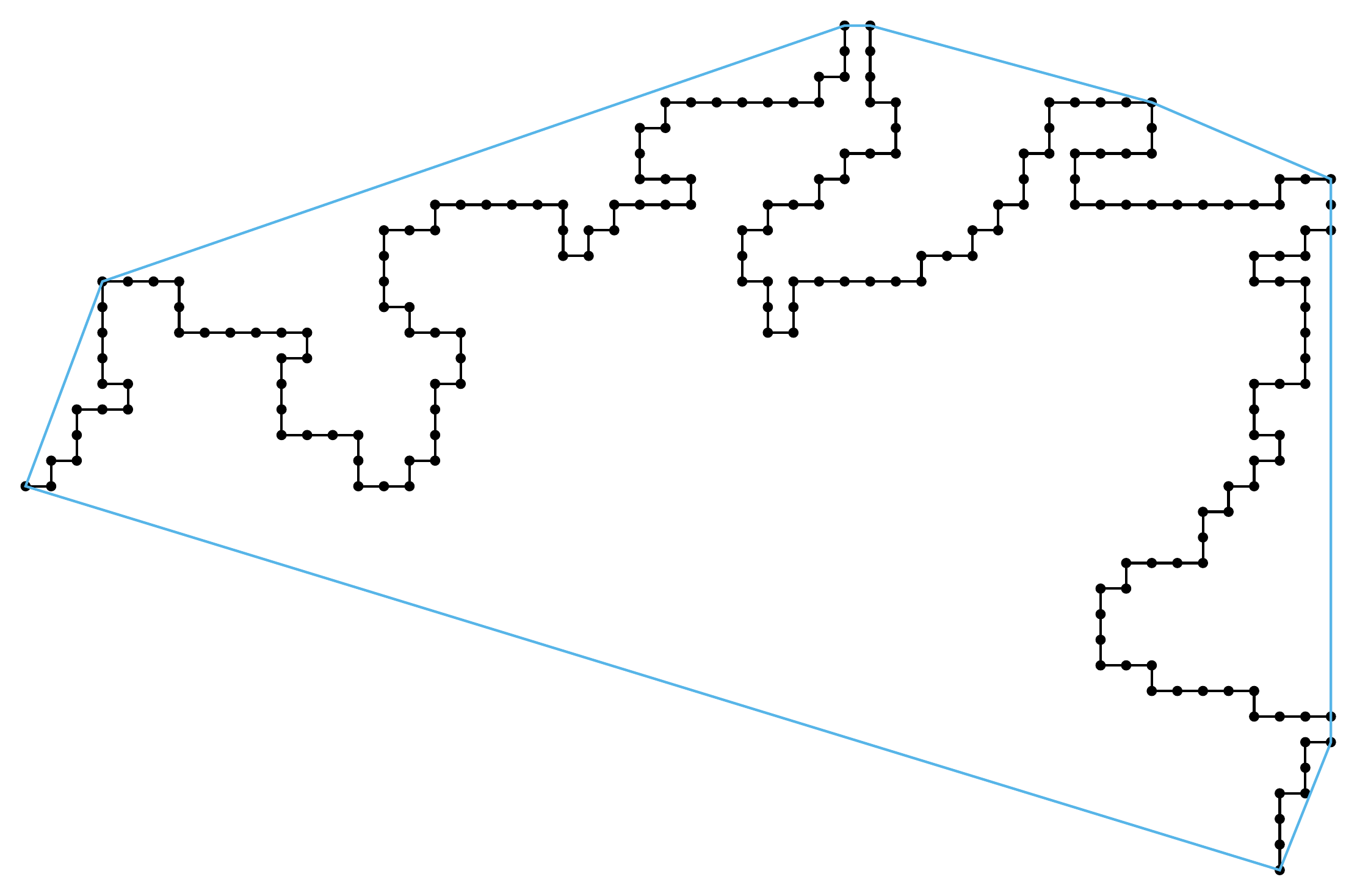}
            }
            \subfigure[\label{fig:LERW} LERW]{
                \includegraphics[scale=0.17]{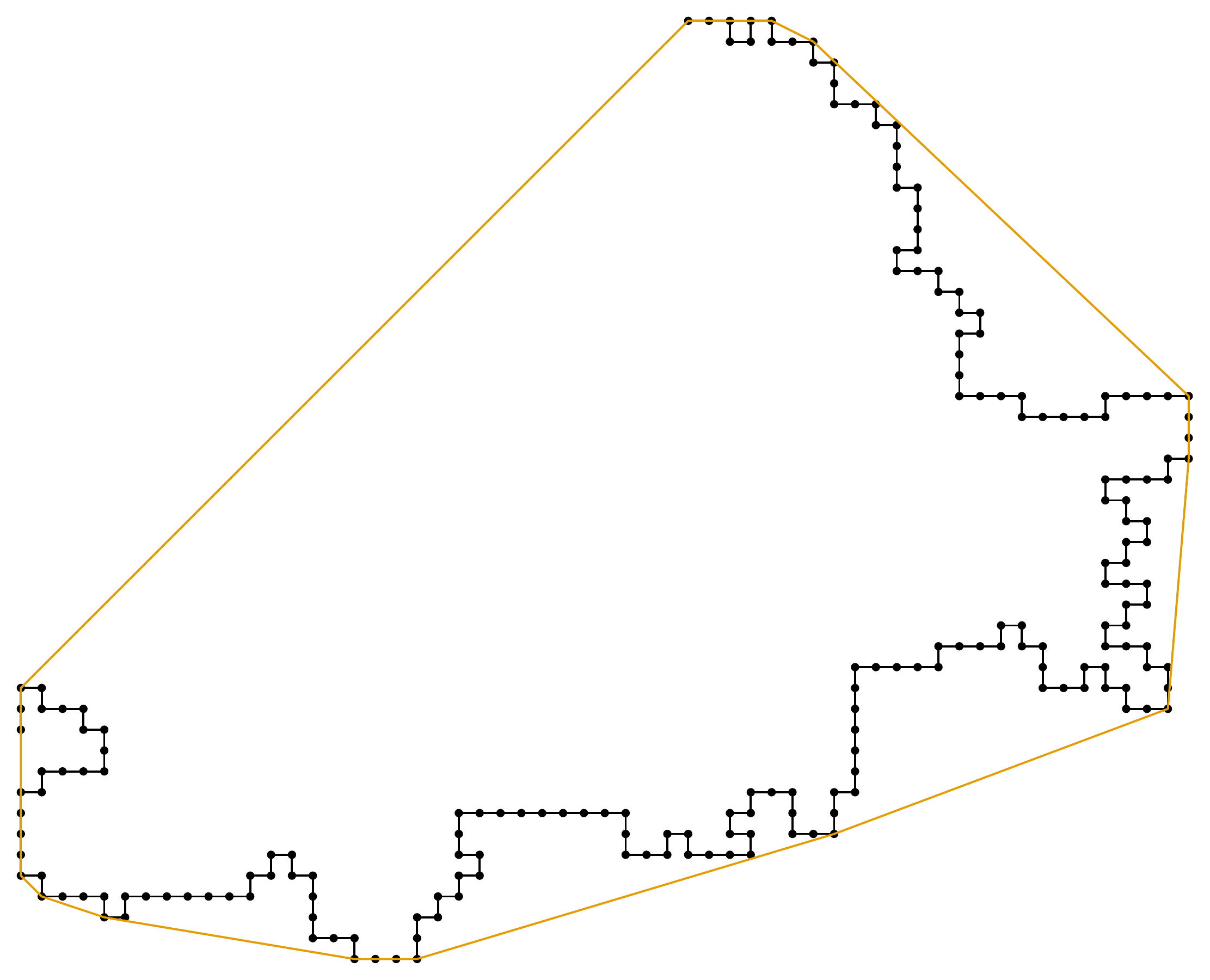}
            }

            \caption{\label{fig:rw}
                (color online)
                Typical configurations with $T=200$ steps, drawn uniformly from
                the corresponding ensembles, of all types of random walks under
                scrutiny in this study with their convex hulls.
            }
        \end{figure}

    \subsection{Smart-Kinetic Self-Avoding Walk (SKSAW)\label{sec:sksaw}}
        The Smart-Kinetic Self-Avoiding
        Walk (SKSAW) \cite{weinrib1985kinetic,kremer1985indefinitely}
        is probably the most naive approach to a self-avoiding walk. It grows
        on a lattice and never enters sites it already visited. Since it is
        possible to get trapped on an island inside already visited sites, this
        walk needs to be \emph{smart} enough to never enter such traps.

        In $d=2$ it is possible to avoid traps using just local information
        in constant time using the \emph{winding angle} method \cite{kremer1985indefinitely}.
        In conjunction with hash table backed detection of occupied sites,
        a realization with $T$ steps can be constructed in time $\mathcal O(T)$.

        This method will typically yield longer stretched walks than the LRW,
        due to the constraint that it needs to be self-avoiding. This can be
        characterized by the exponent $\nu$, which is larger than $1/2$ in $d=2$.

\begin{figure}[bhtp]
            \centering
            \includegraphics[scale=0.8]{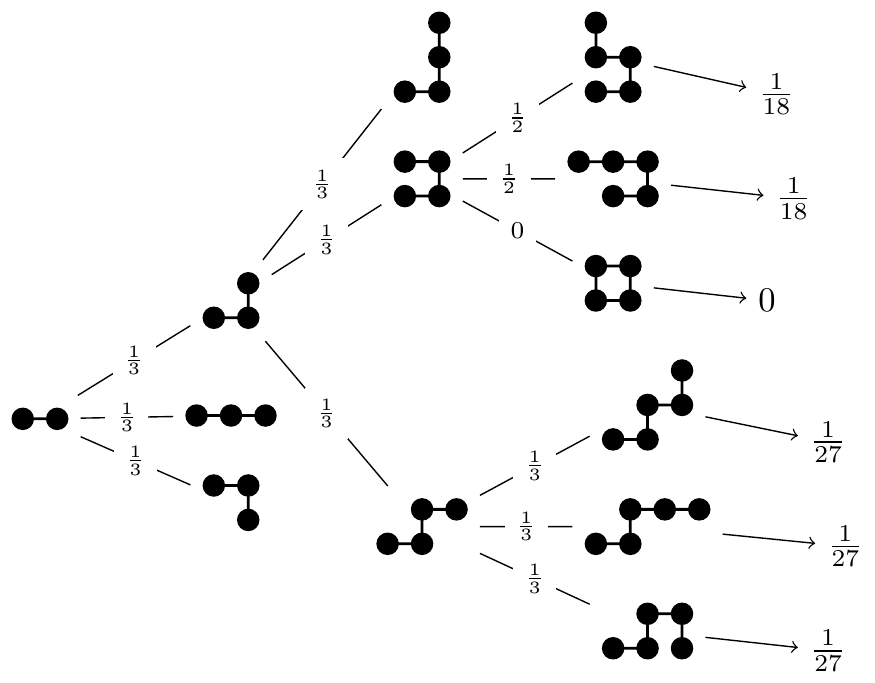}

            \caption{\label{fig:saw_prob}
                Decision tree visualizing the probability to arrive at certain
                configurations following the construction rules of the SKSAW.
                Not all possible configuration have the same probability, hence
                this rules define a different ensemble than SAW.
            }
        \end{figure}

        The sketch Fig.~\ref{fig:saw_prob} shows that this ensemble does not
        contain every configuration with the same probability but prefers
        closely winded configurations. This is also visible in Fig.~\ref{fig:SKSAW}.
        This is characterized by the exponent $\nu=4/7$ \cite{Kennedy2015} which
        is larger than the $\nu$ for LRW, but smaller than for the SAW. Also note
        that it is conjectured that the upper critical dimension is $d=3$ \cite{kremer1985indefinitely}, i.e.,
        $\nu = 1/2$ for all $d \ge 3$ -- possibly with logarithmic corrections
        in $d=3$. Therefore only $d=2$ is simulated in this study.

        While it is easy to draw realizations from this ensemble uniformly, i.e.,
        simple sampling, it is not so straight forward to apply the MCMC changes.
        If one just changes single steps like for the LRW,
        and accepts if it is self-avoiding or rejects if it is not, one will
        generate all self-avoiding walk configurations with equal probability.
        Our approach to generate realizations according to this ensemble handles
        the construction of the walk as a \emph{black box}. It acts on the
        random numbers used to generate a realization from scratch.
        During the MCMC at each iteration one random number is replaced by a new
        random number and a SKSAW realization is regenerated from scratch
        using the modified random numbers \cite{Hartmann2014high}.
        This change is then accepted according to Eq.~\eqref{eq:pacc} and undone
        otherwise.

    \subsection{Self-Avoiding Random Walk (SAW)}
        While the above mentioned SKSAW does produce self-avoiding walks, SAW
        denotes another ensemble. The ensemble where realizations are drawn
        uniformly from the set of all self-avoiding configurations.
        It is not trivial to sample from this distribution efficiently. The
        black box method used for SKSAW is not feasible, since the construction
        of a SAW takes time exponential in the length with simple methods like
        dimerization \cite{dimerization,Madras2013}. It is possible to perform
        changes directly on the walk configuration and accept them according to
        Eq.~\eqref{eq:pacc}, but their rejection rate is typically quite high
        and the resulting configurations are very similar \cite{Madras2013},
        which makes this inefficient. The state of the art
        method to sample SAW is the \emph{pivot algorithm} \cite{Madras2013}.
        It chooses a random point and uses it as the pivot for a random symmetry
        operation, i.e., rotation or mirroring. If the resulting configuration
        is not self avoiding, it is rejected. Otherwise we accept it with the
        temperature dependent acceptance probability Eq.~\eqref{eq:pacc}.

        As mentioned previously, the exponent $\nu=3/4$ \cite{lawler2002scaling} is larger than for the
        SKSAW. Since the upper critical dimension for SAW is $d=4$, this study
        will also look at $d=3$, where an exact value of $\nu$ is not known
        and the best estimate is $\nu = 0.587597(7)$ \cite{Clisby2010Accurate},
        though our focus is on $d=2$ for this type.

        While there are highly efficient implementations of the pivot algorithm
        \cite{Clisby2010Accurate,clisby2010efficient} the time complexity of
        the problem at hand is dominated by the time needed to construct the
        convex hull, thus we go with the simple hash table based
        $\mathcal O(T)$ approach \cite{Madras2013}.

    \subsection{Loop-Erased Random Walk (LERW)}
        The LERW \cite{Lawler1980Self} uses a different approach to achieve the
        self-avoiding property.
        It is built as a simple LRW but each time a site is entered for the
        second time, the loop that is formed, i.e., all steps since the first
        entering of this site, is erased. While this ensures no crossings in
        the walk, the resulting ensemble is different from the SAW ensemble and
        the walks are longer stretched out, as characterized by the larger
        exponent $\nu = 4/5$ \cite{Lawler2011,Guttmann1990Critical,Majumdar1992Exact}. Similar to the SAW
        the upper critical dimension is $d=4$ and an estimate for $d=3$ is
        $\nu = 0.61576(2)$ \cite{Wilson2010}.

        For construction -- similar to SKSAW -- we need to keep all used random
        numbers and change them in the MCMC algorithm. This leads to a
        dramatically higher memory consumption than simple sampling, where
        each loop can be discarded as soon as it is closed.

    \subsection{Convex Hulls}
        We will study the \emph{convex hulls} $\mathcal C$ of the sites visited
        by the random walk $\mathcal P$. The convex hull of a point set $\mathcal P$
        is the smallest polytope containing all Points $P_i \in \mathcal P$ and all
        line segments $(P_i, P_j)$. Some example hulls are shown in Fig.~\ref{fig:rw}.

        Convex hulls are one of the most basic concepts in computational geometry
            \footnote{3 of the first 4 examples for static problems of
                      computational geometry in the Wikipedia can utilize convex
                      hulls for their solution (\protect\url{https://en.wikipedia.org/wiki/Computational_geometry}, 12.01.2018).}
        with noteworthy application in the construction of
        Voronoi diagrams and Delaunay triangulations \cite{brown1979Voronoi}.

        For point sets in the $d=2$ plane, we use Andrew's \emph{Monotone Chain} \cite{Andrew1979Another}
        algorithm for its simplicity and \emph{Quickhull} \cite{Bykat1978Convex} as
        implemented by \emph{qhull} \cite{Barber1996thequickhull} for $d=3$. Both
        algorithms have a time complexity of $\mathcal O(T \ln T)$.
        In $d=2$ Andrew's Monotone Chain algorithm results in ordered points
        of the convex hull. Adjacent points $(i, j)$ in this ordering are the line
        segments of the convex hull. Quickhull results in the simplical facets
        of the convex hull.

        To obtain the perimeter of a $d=2$ convex hull, we sum the lengths of
        its line segments $L_{ij}$.
        To calculate the area and the volume, we use the same fundamental idea.
        In both cases we subdivide the area/volume into simplexes, i.e.,
        triangles for the area and tetrahedra for the volume. Therefore we
        choose an arbitrary fixed point $p_0$ inside of the convex hull and
        construct a simplex for each facet $f_m$, i.e., for $d=2$ each line segment
        of the hull $f_m = (i, j)$ forms a triangle $(i, j, p_0)$ and each
        triangular face $f_m = (i, j, k)$ of a $d=3$ dimensional polyhedron, forms a
        tetrahedron with $p_0$. The volume of a triangle is trivially
        \begin{align*}
            A_{ijp_0} = \frac{1}{2} \dist(f_m, p_0) L_{ij},
        \end{align*}
         where $\dist(f_m, p_0)$ is
        the perpendicular distance from a facet $f_m$ to a point $p_0$.
        Since the union of all triangles built this way, is the whole polygon,
        the sum of their areas is the area of the polygon.
        Similar the volume of a polyhedron is the sum of the volumes of all
        tetrahedra constructed from its faces. The volume of the individual
        tetrahedra is given by
        \begin{align*}
            V_{ijkp_0} = \frac{1}{3} \dist(f_m, p_0) A_{ijk}.
        \end{align*}

        For random walks on a lattice with $T$ steps of length $1$ in $d$ dimensions
        the maximum volume is
        \begin{align}
            \label{eq:max}
            S_\mathrm{max} = \frac{(T/\deff)^\deff}{\deff!}
        \end{align}
        for $T$ divisible by the effective dimension $\deff$ of the
        observable, e.g., 2 for the area of a planar hull or 3 for the volume
        in three dimensions. For example, the configuration of maximum area
        corresponds to an L-shape, i.e., $A_\mathrm{max} = \frac{T^2}{8}$.
        This form can be derived by the general volume of an
        $d$-dimensional simplex defined by its $d+1$ vertices $\vec{v}_i$ \cite{Stein1966Volume}
        \begin{align}
            V = \frac{1}{d!} \det{(\vec{v}_1 - \vec{v}_0, \ldots, \vec{v}_d - \vec{v}_0)}.
        \end{align}
        Without loss of generality, we set $\vec{v}_0$ to be the coordinate
        origin. To achieve maximum volume all $\vec{v}_i, i>0$ need to be
        orthogonal and of equal length. Thus a random walk going $T/d$
        steps along some base vector $\vec{e}_i$ and continuing with $T/d$ steps
        in direction $\vec{e}_{i+1}$ has a convex hull defined by the tetrahedron
        specified by $\vec{v}_i = \sum_{j=1}^{i} \frac{T}{d} \vec{e}_j$.
        The matrix $M = (\vec{v}_1, \ldots, \vec{v}_d)$ is thus triangular and its
        determinant is the product of its diagonal entries $M_{ii} = \frac{T}{d}$
        which leads directly to Eq.~\eqref{eq:max}.
        An exception occurs in $d=2$ where the perimeter is $L_\mathrm{max} = 2T$.

    \section{Results}
        The focus of this work lies on $d=2$ dimensional SAW and LERW. The
        results for higher dimensions and for SKSAW are generated with less
        numerical accuracy. The LRW results also have a lower accuracy as their
        purpose is mainly to scrutinize the effect of the lattice structure
        underlying all considered walk types in comparison to the non-lattice
        results from \cite{schawe2017highdim}. Also not all combinations are
        simulated, but only those listed with a value in Table~\ref{tab:kappa}.

        The same raw data is evaluated for equidistant bins and logarithmic
        bins. And the respective variants are shown according to the scaling of
        the $x$-axis.

    \subsection{Correlations}
\begin{figure}[bhtp]
            \centering
            \includegraphics[scale=1]{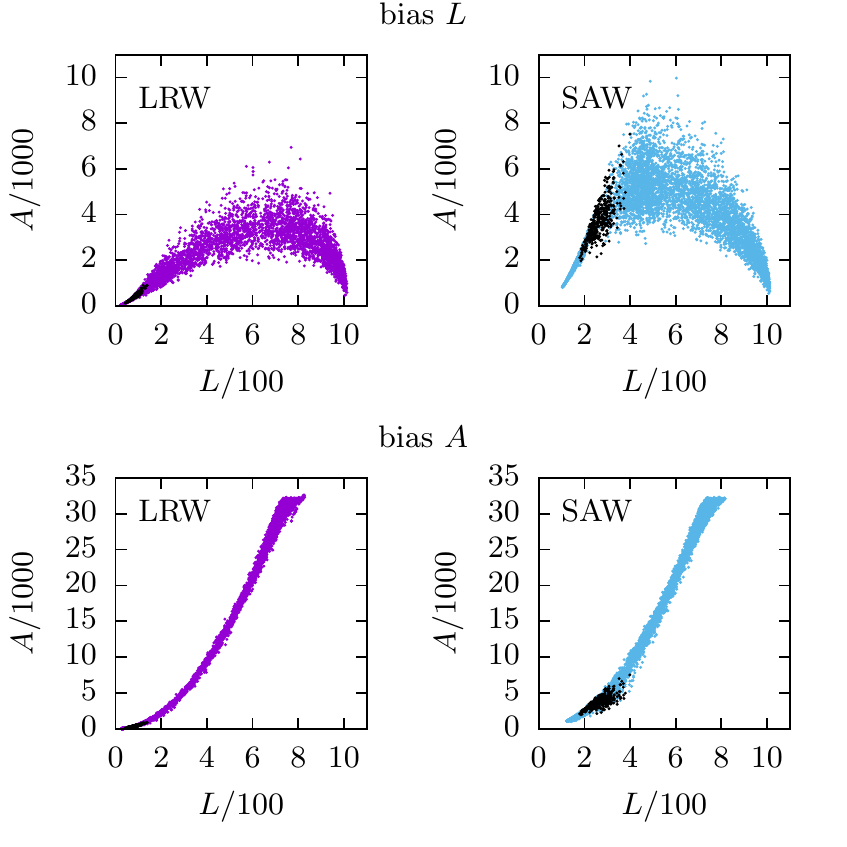}
            \caption{\label{fig:scatter}
                (color online)
                The top row shows data from simulations biasing towards larger
                (and smaller) than typical perimeters $L$. The bottom row biases
                the area $A$. The left column shows data from LRW and the right
                from SAW both with $T=512$ steps. The results of simple sampling
                are shown in black. Note that only very narrow parts are
                covered by simple sampling for the LRW.
            }
        \end{figure}

        To get an intuition for how the configurations with atypical large areas $A$
        or perimeters $L$ look like, we visualize the correlation between these
        two observables as scatter plots in Fig.~\ref{fig:scatter}.

        Since the smallest possible SAW is an (almost) fully filled square,
        there can not be instances below some threshold, which explains the
        gaps on the left side of the scatter plots and of the distributions
        shown in the following section.
        In the center of the scatter plots, which is already in probability
        regions far beyond the capabilities of simple sampling methods, the
        behavior becomes strongly dependent on the bias.

        If biasing for large perimeters (top) the area shows a non-monotonous
        behavior. First, somehow larger perimeters come along typically
        with larger areas for entropic
        reasons, i.e., there are less configurations which are long and thin,
        and more bulky, which have a larger area.
        Though, for the far right tail, the only configurations with extreme
        large perimeters are almost line like and have thus a very small area.
        Also note that the excluded volume effect of the SAW leads to overall
        larger areas at the same perimeters.

        On the other hand, when biasing for large areas (bottom) the
        configurations with largest area, which are
        L-shaped (cf.~Fig.~\ref{fig:saw_temp_cmp}), unavoidably have quite
        large perimeters, hence the scatter plots show an almost linear
        correlation between area and perimeter. Since the configurations of
        large areas naturally avoid self intersections, since steps on already
        visited points do not enlarge the convex hull, the differences between
        LRW and SAW diminish in the right tail. Note that with the
        large-area bias, no walks with the very extreme perimeters exist, for
        the reason already mentioned.

        Note however that these scatter plots are very dependent on which
        observable we are biasing for. In principle we observe
        that small perimeters are strongly correlated with small areas
        while for large but not too large perimeters, there is a broad
        range of area sizes possible. For extremely large perimeters,
        the area must be small.
        For a comprehensive analysis, one would need a full two dimensional
        histogram, wich could be obtained using Wang Landau sampling, but which
        is beyond the scope of this study and would require
        a much larger numerical effort. Nevertheless, from looking at
        Fig.~\ref{fig:scatter} one can anticipate that the two dimensional
        histogram would exhibit a strong correlation for small values of $L$
        and a broad scatter of the accessible values of $A$ for larger but not
        too large  values of $L$.

    \subsection{Moments and Distributions}
\begin{figure}[bhtp]
            \centering
            \includegraphics[scale=1]{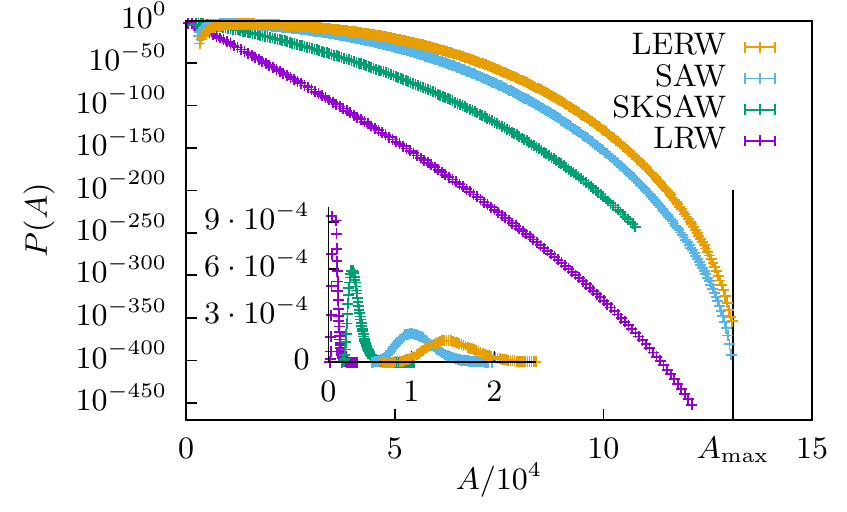}
            \caption{\label{fig:compare}
                (color online)
                Distribution of all scrutinized walk types with $T=1024$ steps.
                The vertical line at $A_\mathrm{max}=131072$ denotes the maximum area (Eq.~\eqref{eq:max}),
                i.e., SAW and LERW are sampled across their full support and
                SKSAW and LRW are not.
                The inset shows the peak region. The gap on the left is due to
                excluded volume effects, i.e., there are no configurations with
                area below some threshold, since this would require
                self-intersection.
            }
        \end{figure}

        The distributions of the different walk types differ considerably. This
        can be observed in Fig.~\ref{fig:compare}, where distributions of the
        area $A$ for all types with $T=1024$ steps are drawn.
        The main part of the distribution shifts to larger values for larger
        value of $\nu$ as expected and the probability of atypically large
        areas is boosted even more in the tails.

        In the right tail, the distributions seem to bend down. Below,
        where we show results for different walk sizes $T$, we see that this
        is a finite-size effect of the lattice structure and the fixed
        step length. This can be seen also as follows:
        Since the lattice together with the fixed step length
        sets an upper bound on the area, the probability plummets near this
        bound for entropic reasons, i.e., there are for any walk length $T$
        only 8 configurations with
        maximum area (due to symmetries) such that all self-avoiding types
        will meet at this point. (not visible because the bins are not fine enough)

        This is supported from Ref.~\cite{Claussen2015Convex} which
        shows that the distribution $P(A)$ for standard random walks with Gaussian jumps,
        i.e., without lattice or fixed step length, do not bend down and have an
        exponential right tail. We conclude that the deviation from this are
        thus caused by this difference.

        First we will look at the rescaled means $\mu_S = \duck{S} / T^{\deff\nu}$,
        where $S$ is an observable and $\deff$ its effective dimension, as
        introduced above in Eq.~\eqref{eq:max}.
        The scaling is a combination of the scaling of the end-to-end distance
        $r \propto T^\nu$ and the typical scaling that a $d$-dimensional
        observable scales as $r^d$ with a characteristic length $r$.

        \begin{figure}[bhtp]
            \centering
            \includegraphics[scale=1]{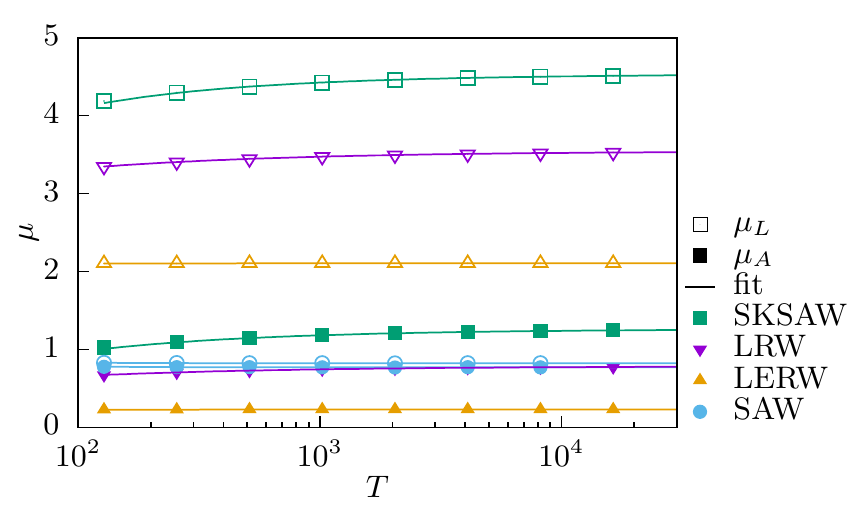}
            \caption{\label{fig:means}
                (color online)
                Scaled means $\mu_A = \duck{A} / T^{2\nu}$ and $\mu_L = \duck{L} / T^{\nu}$
                for different walk types. The lines are fits to extrapolate the
                asymptotic values shown in Table~\ref{tab:measuredMu}
                according to Eq.~\eqref{eq:extrapolation}. Errorbars of the
                values are smaller than the line of the fit and not shown for
                clarity.
            }
        \end{figure}

        Nevertheless, due to finite-size corrections, the ratios
        $\mu_S=\duck{S} / T^{\deff\nu}$ will still depend on the walk length. Thus,
        the measured estimates
        $\mu_S=\mu_S(T)$ at specific walk lengths $T$ need to be extrapolated
        to get an estimate of the asymptotic value
        $\mu_S^\infty = \lim_{T\to \infty}\mu_S(T)$.
        For the extrapolation we use \cite{schawe2017highdim}
        \begin{align}
            \label{eq:extrapolation}
            \mu_S(T) = \mu_S^\infty + C_1 T^{-1/2} + C_2 T^{-1} + o(T^{-1}).
        \end{align}
        This choice is motivated by a large $T$ expansion for the area $A$
        ($\deff=2$) of the convex hulls of standard random walks ($\nu=1/2$)
        with Gaussian jumps \cite{grebenkov2017mean}
        \begin{align}
            \frac{\duck{A}}{T} = \frac{\pi}{2} + \gamma \sqrt{8\pi}\, T^{-1/2} + \pi(1/4+\gamma^2)\, T^{-1}+ o(T^{-1}),
        \end{align}
        where the constant $\gamma= \zeta(1/2)/\sqrt{2\pi}=-0.58259\dots$.
        A natural guess for a generalization to oberservables of a different
        effective dimension $\deff$ \cite{schawe2017highdim} and different
        walk types would be a similar behavior with different coefficients like
        Eq.~\eqref{eq:extrapolation}.

        Indeed, using this form to estimate the asymptotic means $\mu_S^\infty$
        of the observable $S$ yields good fits, as visible in
        Fig.~\ref{fig:means}. In fact, for the fit quality we obtain
        $\chi_\mathrm{red}^2$ values between $0.4$ and $1.7$ (the fit ranges
        for SKSAW begin at $T=512$, for LRW, SAW and LERW at $T=128$, hinting at
        more severe corrections to scaling for the former).
        We assume that the scaling is thus valid for arbitrary random walk types.
        The resulting fit parameters are shown in Table~\ref{tab:measuredMu}.

        For standard random walks with Gaussian jumps the asymptotic means
        $\mu^\infty_{S,\mathrm{Gaussian}}$ are known \cite{Eldan2014Volumetric}.
        These results can be used to predict the corresponding values for LRW.
        First consider the following heuristic argument for a $d=2$ square
        lattice. On average a random walk takes the same amount of
        steps in $x$ and $y$ direction, such that on average two steps displace
        the walker by $\sqrt{2}$, i.e., the diagonal of a square. In contrast a
        Gaussian walker with variance $1$ will be displaced on average by
        $1$ every step. To make both types comparable, we can increase the
        lattice constant to $\sqrt{2}$, which leads to an average displacement
        of $1$ per step for the LRW. Using the same argumentation for higher
        dimensions, we  can use the trivial scaling with the lattice constant
        $S^\deff$ and the length of the diagonal of a unit hypercube $d^{1/2}$,
        to derive a general conversion:
        \begin{align}
            \label{eq:lattice}
            \mu_{S,\mathrm{LRW}}^\infty = \mu_{S,\mathrm{Gaussian}}^\infty / d^{\deff / 2}.
        \end{align}
        These known results are listed next to our measurements in Table~\ref{tab:measuredMu}
        and are within errorbars compatible with our measurements.

        \begin{table}[htb]
            \begin{ruledtabular}
                \begin{tabular}{rlllll}
                     & \multicolumn{1}{c}{$\mu_L^\infty$} & \multicolumn{1}{c}{$\mu_A^\infty$} & \multicolumn{1}{c}{$\mu_{\partial V}^\infty$} & \multicolumn{1}{c}{$\mu_V^\infty$}\\[0.05cm]
                    \hline
                    \noalign{\vskip 0.1cm}
                    LRW (exact) & $3.5449...$  & $0.7854...$ & $2.0944...$ & $0.21440...$\\
                    LRW         & $3.5441(7)$  & $0.7852(2)$ & $2.0945(4)$ & $0.21445(4)$\\
                    SKSAW       & $4.5355(12)$ & $1.2642(5)$ & - & -\\
                    SAW         & $0.8233(7)$  & $0.7714(1)$ & $2.069(2)$  & $0.1996(2)$\\
                    LERW        & $2.1060(3)$  & $0.2300(1)$ & $1.6436(2)$ & $0.13908(3)$\\
                \end{tabular}
            \end{ruledtabular}
            \caption{\label{tab:measuredMu}
                Asymptotic mean values extrapolated from simulational data
                and the exactly known values for the standard random walk (LRW).
                The columns labeled with $\mu_L^\infty$ and $\mu_A^\infty$
                are for $d=2$, those labeled with $\mu_{\partial V}^\infty$ and
                $\mu_V^\infty$ are for $d=3$. For $d=3$ we did not simulate the
                SKSAW, see Section~\ref{sec:sksaw}. Also SAW has lower
                accuracy because of fewer samples in $d=3$.
            }
        \end{table}
        \begin{figure*}[bhtp]
            \centering
            \subfigure[\label{fig:scaling:LRW} LRW]{
                \includegraphics[scale=1]{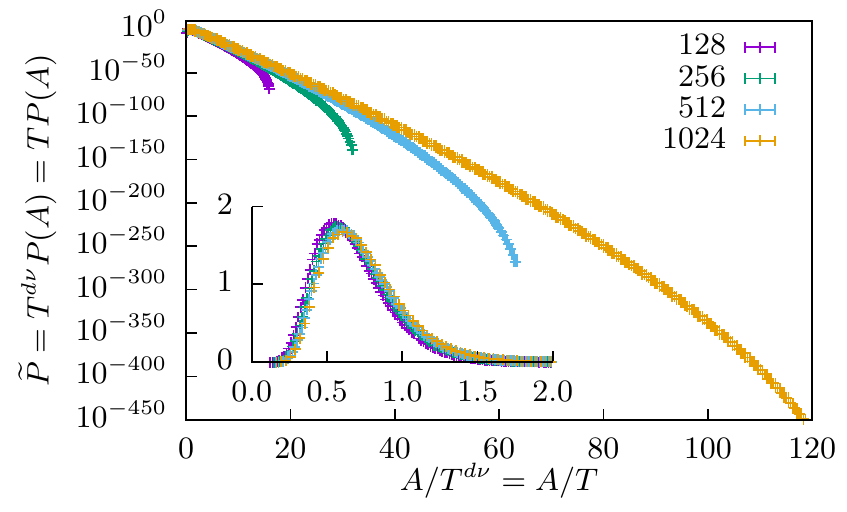}
            }
            \subfigure[\label{fig:scaling:SKSAW} SKSAW]{
                \includegraphics[scale=1]{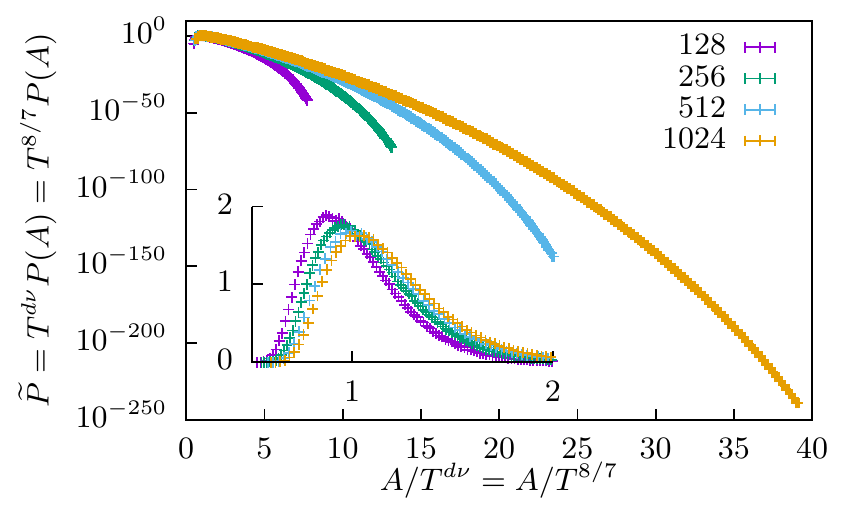}
            }

            \subfigure[\label{fig:scaling:SAW} SAW]{
                \includegraphics[scale=1]{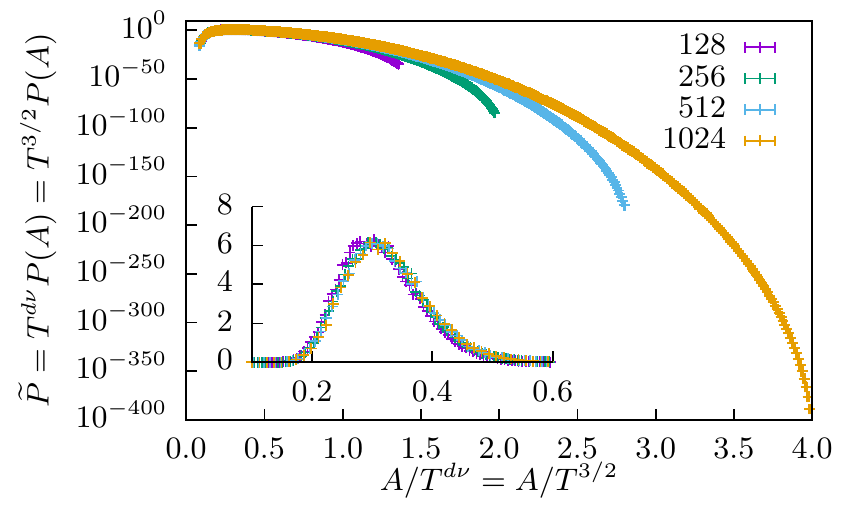}
            }
            \subfigure[\label{fig:scaling:LERW} LERW]{
                \includegraphics[scale=1]{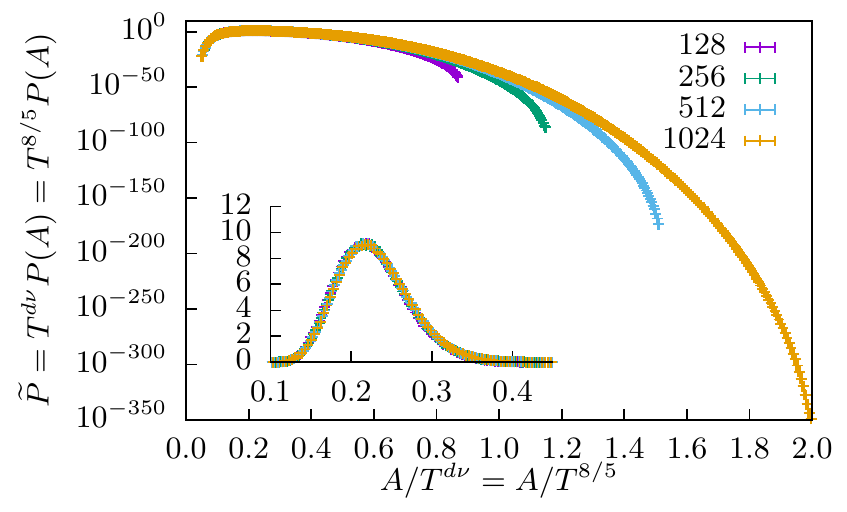}
            }
            \caption{\label{fig:scaling}
                (color online)
                Distributions of the area of different types of random walks
                scaled according to Eq.~\eqref{eq:scaling} for different walk
                lengths $T$.
            }
        \end{figure*}

        Since we have data for the whole distributions, a natural question is,
        whether this scaling does apply over the whole support of the distribution.
        There is evidence that this is true for the convex hulls of standard
        random walks \cite{Claussen2015Convex} in arbitrary dimensions \cite{schawe2017highdim}.
        That means the distributions of an observable $S$ for different
        walk lengths $T$ should collapse onto one universal function
        \begin{align}
            \label{eq:scaling}
            P(S) = T^{-\deff \nu} \widetilde{P}(ST^{-\deff\nu}).
        \end{align}

        Fig.~\ref{fig:scaling} shows the distributions of the $d=2$ area of
        all considered random walk types scaled according to Eq.~\eqref{eq:scaling}.
        The curves collapse well in the peak region and in the intermediate
        right tail. In the far right tail clear deviations from a universal
        curve are obvious, which are the mentioned finite-size effects caused
        by the lattice.

        The distributions look qualitatively similar, though with weaker finite
        size effects, i.e., a better collapse, for the perimeter $L$ (not
        shown). In $d=3$, where we have studied the volume, the results also
        look similar but exhibit stronger finite-size effects (not shown).

\begin{figure*}[bhtp]
            \centering
            \subfigure[\label{fig:rate:LRW} LRW]{
                \includegraphics[scale=1]{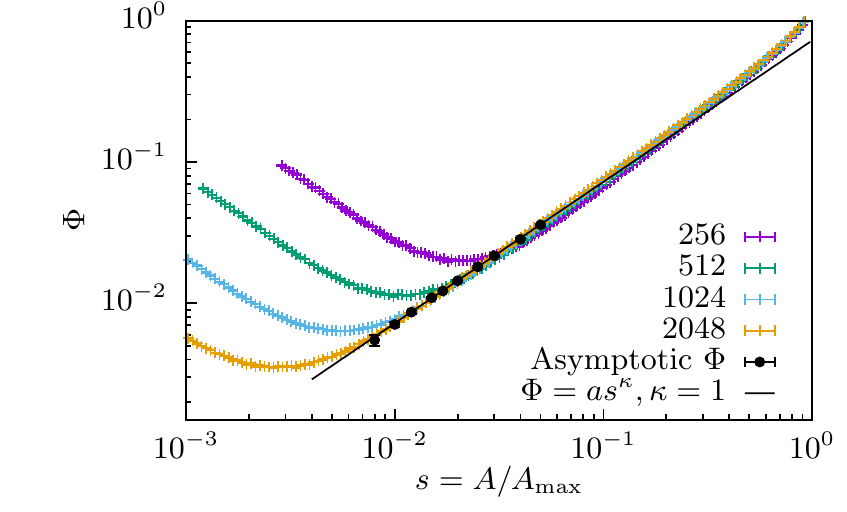}
            }
            \subfigure[\label{fig:rate:SKSAW} SKSAW]{
                \includegraphics[scale=1]{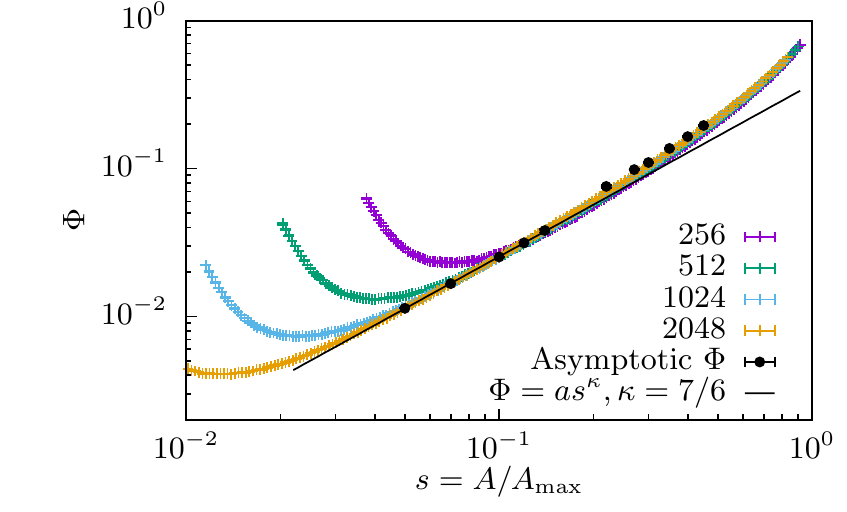}
            }

            \subfigure[\label{fig:rate:SAW} SAW]{
                \includegraphics[scale=1]{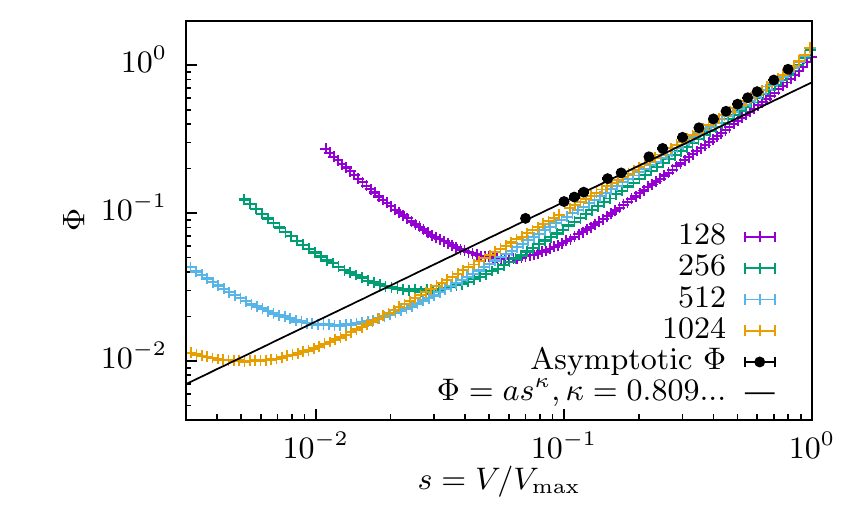}
            }
            \subfigure[\label{fig:rate:LERW} LERW]{
                \includegraphics[scale=1]{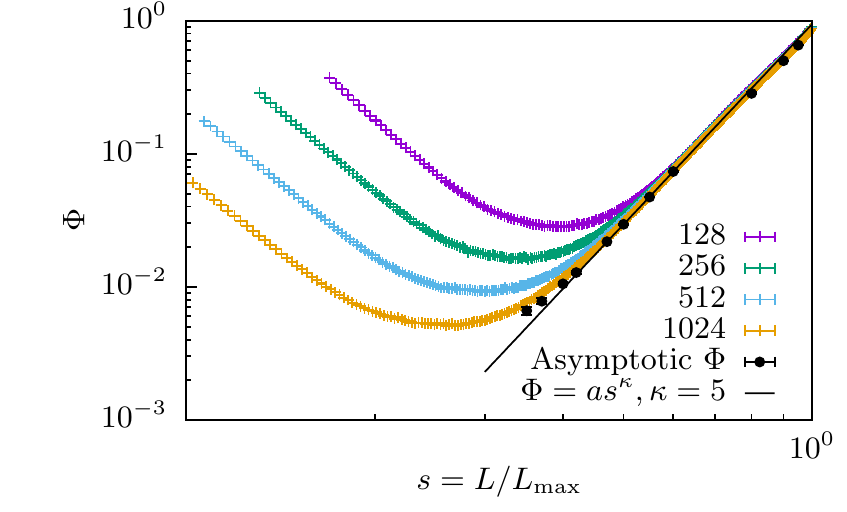}
            }
            \caption{\label{fig:rate}
                (color online)
                Selection of asymptotic rate functions extrapolated from our
                data and our expected exponent $\kappa$ of the rate function $\Phi$.
            }
        \end{figure*}

        Using the full distributions at different values of the walk length $P_T$,
        we can test if it obeys the large deviation principle, i.e., if $\Phi$
        exists, such that the distribution scales as
        \begin{align}
            \label{eq:largeDev}
            P_T \approx \ee^{-T\Phi}
        \end{align}
        for large values of $T$ \cite{Touchette2009large}.
        To simplify comparison, the support of the rate function is usually
        normalized to $[0, 1]$. Here we achieve this by
        using the maximum Eq.~\eqref{eq:max}.
        Solving Eq.~\eqref{eq:largeDev} for $\Phi$ results in
        \begin{align}
            \Phi(S/S_\mathrm{max}) = -\frac{1}{T} \ln P(S/S_\mathrm{max}).
        \end{align}
        We plot this for a selection of our results in Fig.~\ref{fig:rate}.
        From these plots, $\Phi$ seems to approximately follow a power law in
        the intermediate right tail, while the finite-size effects caused by the
        lattice play a major role in the far right tail, which ``bends up''
        consequently.

        Assuming that the rate function behaves approximately as a power law,
        which seems consistent with our data shown in Fig.~\ref{fig:rate}, i.e.,
        \begin{align}
            \Phi(s) \propto s^{\kappa},
            \label{eq:power-law}
        \end{align}
        the exponent $\kappa$ can be estimated by combining the definition of
        $\Phi$ Eq.~\eqref{eq:largeDev} with the scaling assumption
        Eq.~\eqref{eq:scaling} as follows, note that for clarity we use here
        $S_\mathrm{max} \propto T^{\deff}$.
        \begin{align}
            \exp\brac{-T \Phi(S/T^{\deff})}  \sim \frac{1}{T^{\nu \deff}} \widetilde{P}(S/T^{\nu \deff})
        \end{align}
        The $1/T^{\nu \deff}$ term on the right-hand
        side can be ignored next to the exponential. Since the right-hand side
        is a function of $S/T^{\nu \deff}$, the left-hand side must
        also be only dependent on $S/T^{\nu \deff}$.
        This can be achieved by assuming
        $-\nu \deff \kappa + \deff \kappa = 1$, as one can easily see
        by using Eq.~(\ref{eq:power-law}):
        \begin{align*}
            \intertext{Starting from the left-hand side}
            & \exp\brac{-T^{1} \Phi(S/T^{\deff})} \\
                                        \propto& \exp\brac{-T^{1} \brac{S/T^{\deff}}^{\kappa}}\\
                                        =& \exp\brac{-T^{-\nu \deff \kappa + \deff \kappa} \brac{S/T^{\deff}}^{\kappa}}\\
                                        =& \exp\brac{-\brac{S/T^{\nu \deff}}^{\kappa}}
        \end{align*}
        From this we can conclude
        \begin{align}
            \label{eq:kappa}
            \kappa = \frac{1}{d_{\mathrm{e}}(1-\nu)},
        \end{align}
        which simplifies to the case of the standard random
        walk above the critical dimension of the given walk type \cite{schawe2017highdim}
        \begin{align*}
            \kappa_g = \frac{2}{d_{\mathrm{e}}}.
        \end{align*}

        To compare this crude estimate with the results of our simulations,
        we do a point-wise extrapolation of the empirical rate functions for
        fixed walk lengths $T$ as done before in~\cite{Claussen2015Convex,Dewenter2016Convex,schawe2017highdim}.
        For the point-wise extrapolation, we use measurements $\Phi_T$ for
        multiple values of the walk length $T$ at fixed values of
        $S/S_\mathrm{max}$. Since our data are discrete due to binning, the
        values of $\Phi_T$ are obtained by cubic spline interpolation. With
        these data points, which can be thought of as vertical slices through
        the plots of Fig.~\ref{fig:rate}, we extrapolate the $T\to\infty$ case
        with a fit to a power law with offset
        \begin{align}
            \Phi = a T^b + \Phi_\infty.
        \end{align}
        The extrapolated values are marked with black dots in Fig.~\ref{fig:rate}.
        Since finite-size effects have major impact on the tails due to the
        lattice structure, we expect that our estimate is only valid for
        the intermediate right tail of our simulational data. To estimate
        sensible uncertainties, we fit different ranges of our data and give
        the center of the range of the obtained $\kappa$ as our estimate with an
        error including the extremes of the obtained $\kappa$. The black
        lines in Fig.~\ref{fig:rate} are our expected values, which are in all
        examples compatible with some range of our extrapolated data.

        All exponents $\kappa$ we calculated, together with our
        expectations, are listed in Table~\ref{tab:kappa}. A more detailed
        discussion of the examples shown in Fig.~\ref{fig:rate} follows.

        In Fig.~\ref{fig:rate:LRW} the LRW is shown, which is equivalent to Brownian
        motion in the large $T$ limit for which Ref.~\cite{Claussen2015Convex,schawe2017highdim}
        showed the rate function to behave like a power law with exponent $\kappa=1$
        for the area in $d=2$. Using the above mentioned procedure we obtain
        $\kappa = 0.99(2)$ which is in perfect agreement with the
        expectation $\kappa = 1$.

        Fig.~\ref{fig:rate:SKSAW} shows the same for the SKSAW. The obtained
        asymptotic rate function's exponent $\kappa = 1.28(12)$ is compatible
        with our expectation, though the stronger finite-size effects, lead to
        larger uncertainties of our estimate.

        Fig.~\ref{fig:rate:SAW} shows the same but for the volume of the SAW in
        $d=3$ dimensions. The finite-size effects are apparently stronger for
        the volume in $d=3$, as the slope of the right-tail rate function
        gets less steep with increasing system size.

        Fig.~\ref{fig:rate:LERW} for the perimeter of a $d=2$ dimensional LERW.
        In contrast to the area and volume the far right tail of the perimeter
        seems to bend down instead of up, albeit slightly. Though in the
        intermediate right tail, the rate function seems to behave as expected.

        \begin{table}[htb]
            \centering
            \begin{ruledtabular}
                \begin{tabular}{l r l r l}
                    \multirow{ 2}{*}{} & \multicolumn{2}{c}{$V$} & \multicolumn{2}{c}{$\partial V$}\\
                    \cline{2-3} \cline{4-5}
                    \noalign{\vskip 0.1cm}
                        & \multicolumn{1}{c}{Eq.~\eqref{eq:kappa}} & \multicolumn{1}{c}{$\kappa$} & \multicolumn{1}{c}{Eq.~\eqref{eq:kappa}} & \multicolumn{1}{c}{$\kappa$}\\[0.1cm]
                    LRW        & $1          $ & $0.99(2)$  & $2          $ & -          \\
                    SKSAW      & $\frac{7}{6}$ & $1.28(12)$ & $\frac{7}{3}$ & -          \\
                    SAW        & $2          $ & $2.2(4)$   & $4          $ & $4.11(14)$  \\
                    SAW $d=3$  & $0.809...   $ & $0.92(11)$ & $1.214...   $ & -          \\
                    LERW       & $\frac{5}{2}$ & $2.57(24)$ & $5          $ & $4.82(19)$ \\
                    LERW $d=3$ & $0.867...   $ & $0.89(9)$  & $1.299...   $ & -          \\
                \end{tabular}
            \end{ruledtabular}
            \caption{
                Comparison of expected and measured rate function exponent $\kappa$.
                The value is the center of multiple fit ranges and the error is
                chosen such that the largest and the smallest result is enclosed.
            }
            \label{tab:kappa}
        \end{table}

        In general, our data supports the convergence to a limiting
        rate function, which, mathematically speaking, means that
        the \emph{large-deviation principle} holds. This means that
        the distributions are somehow well behaved and might be
        accessible to analytical calculations. Though the estimate
        for what the rate function $\Phi$ actually is, can possibly be improved.
        However, since our estimate for $\kappa$ is always compatible with our
        measurements it appears plausible that also
        for interacting walks the distribution of the convex hulls is
        governed by the scaling behavior of the end-to-end distance,
        as given by the exponents $\nu$.

    \section{Conclusions}\label{sec:conclusion}
        We numerically studied the area and perimeter of the convex hulls of
        different
        types of self-avoiding random walks in the plane and to a lesser
        degree the volume of their convex hulls in $d=3$ dimensional space.
        By applying sophisticated large-deviation algorithms, we
        calculated the full distributions, down to extremely small
        probabilities like $10^{-400}$. We also obtained corresponding rate
        functions of these observables. Our data support a convergence
        of the rate functions, which means the large-deviation
        principle seems to hold. We observed a generalized
        scaling behavior, which was before established for standard random
        walks. Thus, although the self-avoiding types of walk
        exhibit a more complicated behavior as compared to standard
        random  lattice walks, and although the limiting scaled distributions
        of their convex hull's volume and surface look quite different for the
        various walk cases, in the end
        the convex hull behavior seems to be still governed by the single
        end-to-end distance scaling exponent $\nu$.

        We also observed, rather expectedly,
        that the two observables area and perimter are highly correlated
        for small values. For large but not  too large values of the
        perimeter, many different values of the area are possible,
        but statistically dominated by rather small values of the area.
        Extremly large values of the perimeter are only feasible with
        shrinking area.

        Finally, we gave estimates for the large $T$ asymptotic mean values
        of the mentioned observables. These might be of interest for attempts
        to calculate these values analytically.

        For future studies it could be interesting to look closer into the
        correlations between different observables that we briefly noticed.
        For a more throughout study, it would be useful to obtain full
        two-dimensional histograms.

    \begin{acknowledgments}
        This work was supported by the German Science Foundation (DFG) through
        the grant HA 3169/8-1.
        HS and AKH thank the LPTMS for hospitality and financial support during one and two-month
        visits, respectively, where considerable part of the projects were performed.
        The simulations were performed at the HPC clusters HERO and CARL, both
        located at the University of Oldenburg (Germany) and funded by the DFG
        through its Major Research Instrumentation Programme
        (INST 184/108-1 FUGG and INST 184/157-1 FUGG) and the Ministry of
        Science and Culture (MWK) of the Lower Saxony State.
        We also thank the GWDG G\"ottingen for providing computational resources.
    \end{acknowledgments}

    \bibliography{lit}

\end{document}